\documentclass[%
 reprint,
 amsmath,amssymb,
 aps,
]{revtex4-2}
\usepackage{multirow}
\usepackage{xcolor}

\usepackage{graphicx}
\usepackage{dcolumn}
\usepackage{bm}
\usepackage{adjustbox}

\begin{document}

\title{Invisible hand and arbitrage equilibrium in the self-organizing dynamics of pattern formation in ecological systems}

\author{Venkat Venkatasubramanian}%
\email{venkat@columbia.edu}
\affiliation{Complex Resilient Intelligent Systems Laboratory, Department of Chemical Engineering, Columbia University, New York, NY 10027}

\author{Arun Sankar E M} 
\affiliation{Complex Resilient Intelligent Systems Laboratory, Department of Chemical Engineering, Columbia University, New York, NY 10027}

\author{Abhishek Sivaram}
\affiliation{Department of Chemical and Biochemical Engineering, Technical University of Denmark, Kongens Lyngby, Denmark}

\begin{abstract}
Patterns in ecological systems such as mussel beds have been of considerable interest for a long time. Several physicochemical mechanisms have been proposed for their formation. Here, we propose a novel framework based on economics and game theory. Since mussels are biological agents instinctively driven by the \emph{survival purpose}, we mathematically model this purpose explicitly using a new theoretical framework called \textit{statistical teleodynamics}. We show both analytically and computationally that when every mussel pursues its own self-interest to survive, a stable collective order emerges spontaneously via self-organization. Thus, our mechanism is essentially the same as Adam Smith’s invisible hand in a biological context. Our analysis reveals a new insight that the mussel bed patterns could be the result of \textit{arbitrage equilibrium} in the competition for survival among the mussels.  
\end{abstract}

\keywords{Pattern formation$|$ Invisible hand$|$ Statistical teleodynamics$|$ Ecology$|$ Arbitrage equilibrium$|$ } 

\maketitle

\section{Introduction}
Pattern formation and emergent behavior have always been of great interest in chemical, biological, and ecological domains. This has gained particular attention in ecology in recent years as more detailed empirical data have become available for a variety of systems such as mussel beds \cite{koppel2005scale,liu2012alternativemechanisms,deJager2020patterning} and arid bush lands \cite{klausmeier1999regular,rietkerk2002selforganisation,hardenberg2001diversity,borgogno2009mathematical,koppel2006scale_dependent_inhibition}. Progress in large-scale computational modeling capabilities has also contributed to this interest in important ways. \\

A number of mechanisms have been proposed over the years for the emergence of spontaneous patterns. The Turing reaction-diffusion mechanism \cite{turing1952chemical_basis} has long been a leading contender and has a vast literature \cite{murray2002mathematical,kondo2010reaction,maini2003mouse,zhabotinsky1964bperiodical,castets1990experimental,ouyang1991transition,klausmeier1999regular,rietkerk2002selforganisation,hardenberg2001diversity,borgogno2009mathematical,koppel2006scale_dependent_inhibition,koppel2005scale,liu2012alternativemechanisms}. This was later joined by scale-dependent feedback as an important alternative~\cite{koppel2005scale}.
Recently, another candidate has emerged, which is inspired by the pattern formation theories in chemical engineering and material science. This is the Cahn-Hilliard mechanism, which is based on the principles of thermodynamics, that predicts and explains the emergence of phase separation and pattern formation in materials such as alloys and polymer blends \cite{cabral2018spinodal,han1992phase,fetters1999chain,eitouni2007thermodynamics}. Liu et al. \cite{liu2013CahnHilliard} have shown how the Cahn-Hilliard model could be adapted to explain pattern formation in mussel beds. \\

While these mechanisms have adapted ideas from physics and chemistry, here we propose a new perspective that utilizes concepts and techniques from economics and game theory. Since mussels are biological agents driven by the \textit{purpose} to survive and thrive in challenging environments, modeling this survival goal explicitly is central in our theory. We exploit the concept of \textit{utility} from economics and game theory to capture this survival goal. We mathematically model and analyze the utility-driven emergent behavior of mussels using a new theoretical framework called \textit{statistical teleodynamics} \cite{venkat2015howmuch, venkat2017book}. The result is a mathematically simpler and more universal mechanism that has been successfully applied to other dynamical systems in biology \cite{venkat2022unified, venkat2022garuds}, economics \cite{venkat2015howmuch, venkat2017book, kanbur2020occupational}, and sociology \cite{venkat2022unified} to predict emergent phenomena. Our analysis reveals a new insight that the mussel bed patterns could be the result of 
\textit{arbitrage equilibrium} in the competition among the mussels for survival. Furthermore, a Lyapunov analysis of this self-organizing dynamics also reveals the important result that these mussel bed patterns are \textit{asymptotically stable}. \\

The rest of the paper is organized as follows. After a brief introduction to statistical teleodynamics, we show how the self-organizing behavior of survival purpose-driven mussels can be modeled in our framework. We then discuss the analytical and agent-based simulation results. We conclude with a discussion of the main results and their implications. 

\section{Statistical Teleodynamics, Population Games, and Arbitrage Equilibrium}

Statistical teleodynamics is the natural generalization of statistical thermodynamics for goal-driven agents in active matter. It is a synthesis of the central concepts and techniques of population games theory with those of statistical mechanics towards a unified theory of emergent equilibrium phenomena and pattern formation in active matter. The name comes from the Greek word \textit{telos}, which means goal. Just as the dynamical behavior of gas molecules is driven by thermal agitation (hence, \textit{thermo}dynamics), the dynamics of purposeful agents is driven by the pursuit of their goals and, hence, \textit{teleo}dynamics. \\

The theory of population games is concerned with the prediction of the final outcome(s) of a large population of goal-driven agents. Given a large collection of strategically interacting rational agents, where each agent is trying to decide and execute the best possible course of actions that maximizes the agent's~{\em payoff} or {\em utility} in light of similar strategies executed by the other agents, can we predict what strategies would be executed and what outcomes are likely~\cite{easley2010networks, sandholm2010population}? In particular, one would like to know whether such a game would lead to an equilibrium situation.\\

Population games theory provides an analytical framework for studying the strategic interactions of a large population of agents with the following properties, as described by Sandholm~\cite{sandholm2010population}: (i) The number of agents is large; (ii) Individual agents play a small role - any one agent's behavior has only a small effect on other agents' utility; (iii) Agents interact anonymously -- each agent's utility only depends on opponents' behavior through the distribution of their choices; (iv) The number of roles is finite -- each agent is a member of one of a finite number of populations, and (v) Utilities  are continuous -- this property makes sure that very small changes in aggregate behavior do not lead to large changes in utilities. \\

In games like Prisoner's Dilemma, with a small number of players, the typical game-theoretic analysis proceeds by systematically evaluating all the options every player has, determining their utilities, writing down the utility matrix, identifying the best responses of all players, identifying the dominant strategies (if present) for everyone, and then finally reasoning whether there exists a Nash Equilibrium (or multiple equilibria) or not. However, for population games, given a large number of players, this procedure is not feasible for many reasons -- for instance, a player may not know of all the strategies others are executing (or could execute), and their utilities, for her to determine her best response.\\
    
Fortunately, for some population games, one can identify a single scalar-valued global function, called a {\em potential} ($\phi(\boldsymbol{x})$), that captures the necessary information about the utilities {(where $\boldsymbol{x}$ is the state vector of the system)}. The {\em gradient} of the potential is the \textit{payoff} or \textit{utility} (game theory literature prefers the term payoff, but we prefer utility in this paper to minimize jargon). Such games are called {\em potential games}~\cite{rosenthal1973class,sandholm2010population, easley2010networks, monderer1996potential}. A potential game reaches strategic equilibrium, called \textit{Nash equilibrium}, when the potential $\phi(\boldsymbol{x})$ is maximized. Furthermore, this equilibrium is unique if  $\phi(\boldsymbol{x})$ is strictly concave (i.e., $\partial^2 \phi /\partial^2 x < 0$)~\cite{sandholm2010population}. \\

As noted, in potential games, a player's utility, $h_i$, in state $i$ is the gradient of potential $\phi(\boldsymbol{x})$, i.e.,
\begin{equation}
{h}_i(\boldsymbol{x})\equiv {\partial \phi(\boldsymbol{x})}/{\partial x_i}
\end{equation}
where $x_i=N_i/N$ and $\boldsymbol{x}$ is the population vector. {$N_i$ is the number of agents in state $i$, and $N$ is the total number of agents}. Therefore, by integration (we replace partial derivative with total derivative because ${h}_i(\mathbf{x})$ can be reduced to ${h}_i(x_i)$), we have 
\begin{eqnarray}
\phi(\boldsymbol{x})&=&\sum_{i=1}^n\int {h}_i(\boldsymbol{x}){d}x_i \label{eq:potential}
\end{eqnarray}

where $n$ is the total number of states. \\

To determine the maximum potential, one can use the method of Lagrange multipliers with $L$ as the Lagrangian and $\lambda$ as the Lagrange multiplier for the constraint $\sum_{i=1}^nx_i=1$:

\begin{equation}
L=\phi+\lambda(1-\sum_{i=1}^nx_i)
\label{eq:lagrangian}
\end{equation}\\
If there are other constraints, they can be accommodated similarly~\cite{venkat2015howmuch}.\\

At equilibrium, all agents enjoy the same utility, i.e., $h_i = h^*$. It is an \textit{arbitrage equilibrium} \cite{kanbur2020occupational} where the agents do not have any incentive to switch states anymore as all states provide the same utility $h^*$. In other words, equilibrium is reached when the opportunity for arbitrage, i.e., the ability to increase one’s utility by simply switching to another option or state at no cost, disappears. Thus, the maximization of $\phi$ and $h_i = h^*$ are equivalent when the equilibrium is unique (i.e., $\phi(\boldsymbol{x})$ is strictly concave \cite{sandholm2010population}), and both specify the same outcome, namely, an arbitrage equilibrium. The former stipulates it from the \textit{top-down, system perspective} whereas the latter the \textit{bottom-up, agent} perspective. Thus, this formulation exhibits a \emph{duality} property. 

\section{Mechanism of Pattern Formation in Mussel Beds}

 It is seen empirically that mussels self-organize to form clusters, typically displaying three different stable patterns under different conditions. It has been suggested that such self-organization improves their survival chances against predation and the destabilizing effects of wave stress \cite{liu2013CahnHilliard}. \\
 
 We therefore start with the premise that the mussels behavior is driven by their biological instincts to survive and thrive under challenging conditions. This survival-fitness objective is modeled as utility that mussels try to maximize by their dynamic behavior. They maneuver around in their turbulent environment, exploiting arbitrage opportunities to maximize their survival-fitness. This self-organizing dynamical behavior eventually leads to a stable arbitrage equilibrium and pattern formation, as we show below. This mechanism is essentially the same as Adam Smith's \textit{invisible hand} from economics, but now executed in the context of biology and ecology. \\

 We wish to emphasize that we are not claiming that the mussels pursue the survival goal and strategies rationally. Our point is that the biological survival instincts of mussels cause particular dynamical behaviors that were evolved over millions of years to help them improve their survival chances. Thus, they act in a goal-driven manner \emph{instinctively}, which can be modeled using our framework of pursuing maximum utility or survival fitness.\\

\section{Utility Model and Arbitrage Equilibrium}

Our goal is to identify the fundamental principles and mechanisms of self-organization by survival-driven agents. Towards that, we develop simple models that offer an appropriate coarse-grained description of the system. Unlike atoms and molecules modeled by, for example, the Cahn-Hilliard equation for phase separation,  biological, ecological, and sociological entities do not behave precisely and predictably. Therefore, we have tried deliberately to keep the models as simple as possible, and not be restricted by system-specific details and nuances, without losing key insights and relevance to empirical phenomena\cite{venkat2022unified}.\\

We also wish to stress that the spirit of our modeling is similar to that of the ideal gas or the Ising model in statistical thermodynamics. Just as real molecules are not point-like objects, or devoid of intermolecular interactions, as assumed in the ideal gas model in statistical mechanics, we make similar simplifying assumptions in our mussels model. These can be relaxed to make them more realistic in subsequent refinements like, for example, van der Walls did in thermodynamics. The ideal versions serve as useful starting and reference points to develop more comprehensive models of self-organization in biological and ecological systems.\\

Since the central theme in our theory is survival-driven utility maximization by the mussels jockeying for better positions, we formulate the problem by first defining the \emph{effective utility}, $h_i$, for the mussels in state $i$. The effective utility is the net sum of the \textit{benefits} minus the \textit{costs} for the mussels. Biological agents constantly make benefit-cost trade-offs in their self-actuating dynamical behavior to improve their survival fitness, i.e., the effective utility. This results in a delicate balancing, dynamically, of the benefits of aggregation versus the costs of overcrowding of the agents. In other words, the benefits of cooperation are balanced with the costs of competition. \\

In addition, driven by natural instincts, the agents also balance two competing strategies - \textit{exploitation} and \textit{exploration}. Exploitation takes advantage of the opportunities in the immediate, local, neighborhood of the mussels. On the other hand, exploration examines possibilities outside. This is a widely used strategy in biology. For example, a genetic mutation can be thought of as exploitation, searching locally in the design space, whereas crossover is exploration, searching more globally.\\

We believe that this combination of two main strategies, namely, the benefit-cost trade-offs of cooperation-competition strategy with an exploitation-exploration strategy is a fundamental and universal evolutionary mechanism that is found in most living systems.  Biological evolution tries to exploit the fitness advantages of an ecological niche through small-scale adaptations (such as the specialized beaks of Galapagos finches as Darwin observed), while exploring for bigger fitness gains via large-scale changes in the design. \\

We motivate our model by initially considering a discrete version of the mussel bed, modeled as a large lattice of local neighborhoods or blocks, each with $M$ sites which the mussels can occupy. There are $n$ such blocks, $nM$ sites, and a total of $N$ mussels, with an average mussel density of $\rho_0 = N/(nM)$. The state of a mussel is defined by specifying the block $i$ it is in, and state of the system (i.e., the mussel bed) is defined by specifying the number of mussels, $N_i$, in block $i$, for all blocks ($i \in \{1, \dots, n\}$ ). Let block $i$ also have $V_i$ vacant sites, so  $V_i = M - N_i$. This approach is an extension of our recent model developed for a Schelling game-like scenario \cite{venkat2022unified}. \\

We further formulate the problem by defining the effective utility, $h_i$, for the mussels in block $i$, which the mussels try to maximize by moving to better locations (i.e., other blocks), if possible. The effective utility is the net sum of the benefits minus the costs. A mussel instinctively prefers to have more members in its neighborhood as this aggregation improves its chances of survival against predators and the destabilizing effects of wave stress \cite{liu2013CahnHilliard}.  Thus, this \textit{affinity benefit} term, representing cooperation among mussels,  is proportional to the number of mussels in its neighborhood. We model this as $\alpha N_i$, where $\alpha >0$ is a parameter. \\

\begin{figure}[!ht]
\includegraphics[width=0.5\textwidth]{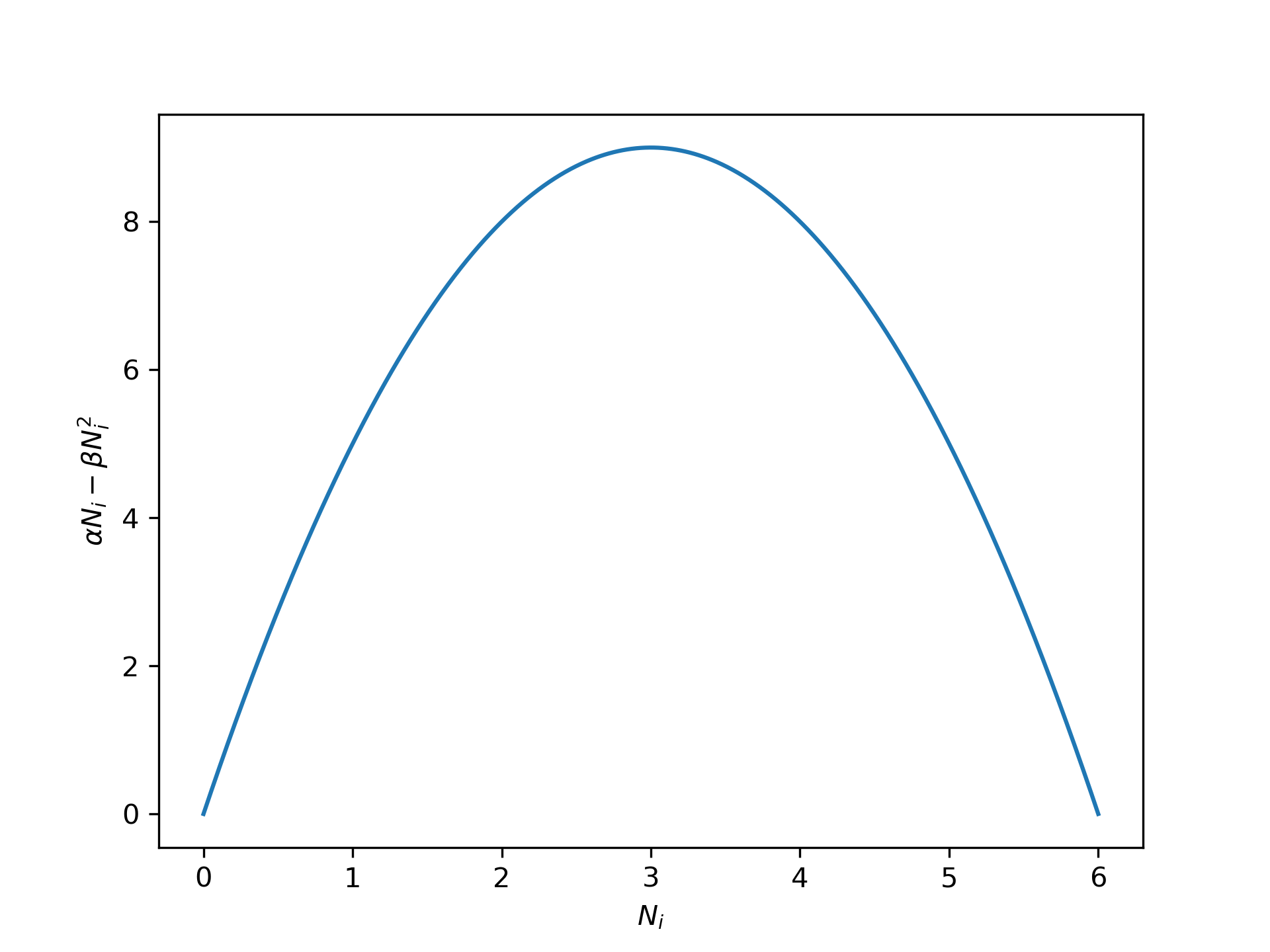}
\caption{Net benefit of a resource for $\alpha N_i - \beta {N_i}^2$ ( $\alpha=6$ , $\beta =1$)}
\label{fig:U_curve}
\end{figure}

However, this affinity benefit comes with a cost.  As more and more mussels aggregate, they all compete for the same and limited food resource in the neighborhood. This is the disutility of overcrowding, and it results in a \textit{congestion cost} term. As Venkatasubramanian explains \cite{venkat2017book}, the resulting  net benefit (= benefit - cost) function has an inverted U-like profile (see Fig.~\ref{fig:U_curve}). This  profile is found in many net benefit vs resource  relationships in the real-world. As one consumes a resource, it initially leads to increasing net benefit; but after a point, the cost of the resource goes up more quickly than the benefit, thus resulting in decreasing net benefit. The simplest model of this is a quadratic function, $\alpha N_i - \beta {N_i}^2$, with the quadratic term $-\beta N_i^2$ ($\beta >0$) modeling the congestion cost.  \\

Regarding exploration, the mussels derive a benefit by having a large number of vacant sites to potentially move to in the future should such a need arise. This is the instinct to explore other opportunities as new vacant sites are potentially new sources of food and other benefits. We call this resource the \textit{option benefit} term as the mussels have the option to move elsewhere if needed. Again, following Venkatasubramanian~\cite{venkat2017book, kanbur2020occupational, venkat2022unified}, we model this as $\gamma \ln (M-N_i)$,  $\gamma >0$. The logarithmic function captures the diminishing utility of this option, a commonly used feature in economics and game theory. As before, this benefit also is associated with a cost due to competition among the mussels for these vacant sites. We model this disutility of competition as $-\delta \ln N_i$, $\delta >0$ \cite{venkat2015howmuch, venkat2017book, kanbur2020occupational}.\\

Combining all these, we have the following effective utility function $h_i$ for the mussels in block $i$ as, 
\begin{eqnarray}
    h_i(N_i) = \alpha N_i  - \beta N_i^2 + \gamma \ln(M - N_i) - \delta \ln N_i
\end{eqnarray}

Intuitively, the first two terms in the equation model the benefit-cost trade-off in the exploitation behavior while the last two model a similar trade-off in exploration. \\

We can set $\delta = 1$ without any loss of generality. In addition, we set $\gamma = 1$ to gain analytical simplicity, but this can be relaxed later if necessary. So we now have

\begin{eqnarray}
    h_i(N_i) = \alpha N_i  - \beta N_i^2 + \ln(M - N_i) - \ln N_i
    \label{eq:Utility_N_i}
\end{eqnarray}

Rewriting this in terms of the density ($\rho_i$) of mussels in block $i$, $\rho_i = N_i/M$, and absorbing the constant $M$ into $\alpha$ and $\beta$, we have

\begin{eqnarray}
    h_i(\boldsymbol \rho) = \alpha \rho_i - \beta \rho_i^2 + \ln (1- \rho_i) - \ln \rho_i
    \label{eq:Utility}
\end{eqnarray}

For the sake of simplicity, we define $u(\rho_i) = \alpha \rho_i - \beta \rho_i^2$. Therefore, the potential $\phi(\boldsymbol{\rho})$ becomes, 

\begin{eqnarray}
    \phi(\boldsymbol{\rho}) &=& \sum_{i=1}^n \int h_i(\boldsymbol{x}) d x_i = \frac{M}{N}\sum_{i=1}^n \int h_i(\boldsymbol{\rho}) d \rho_i \nonumber\\
    &=&\frac{M}{N}\sum_{i=1}^n \int_0^{\rho_i}\left[ u(\rho) + \ln (1-\rho) - \ln \rho \right] d \rho \nonumber\\
    \label{eq:schelling-potential}
\end{eqnarray}
One can generalize the discrete formulation to a continuous one by replacing $\rho_i$ by $\rho(r)$, where the density is a continuous function of radius $r$ of the neighborhood as demonstrated by Sivaram and Venkatasubramanian~\cite{venkat2022garuds} in the self-organized flocking behavior of birds. \\

Now, according to the theory of potential games~\cite{sandholm2010population}, an arbitrage equilibrium is reached when the potential is maximized. We can determine the equilibrium utility, $h^*$, by the Lagrangian multiplier approach mentioned above (\eqref{eq:lagrangian}), but there exists a simpler alternative that is more convenient for our purpose here. To analyze the equilibrium behavior, we can take the simpler agent-based perspective and exploit the fact that at equilibrium all agents have the same effective utility, i.e., $h_i = h^*$, for all $i$. In other words, 

\begin{eqnarray}
        \alpha \rho^* - \beta \rho^{*2} + \ln (1-\rho^*) - \ln\rho^* = h^* 
        \label{eq:utility_equil}
\end{eqnarray}

\begin{figure}[!ht]
\includegraphics[width=0.5\textwidth]{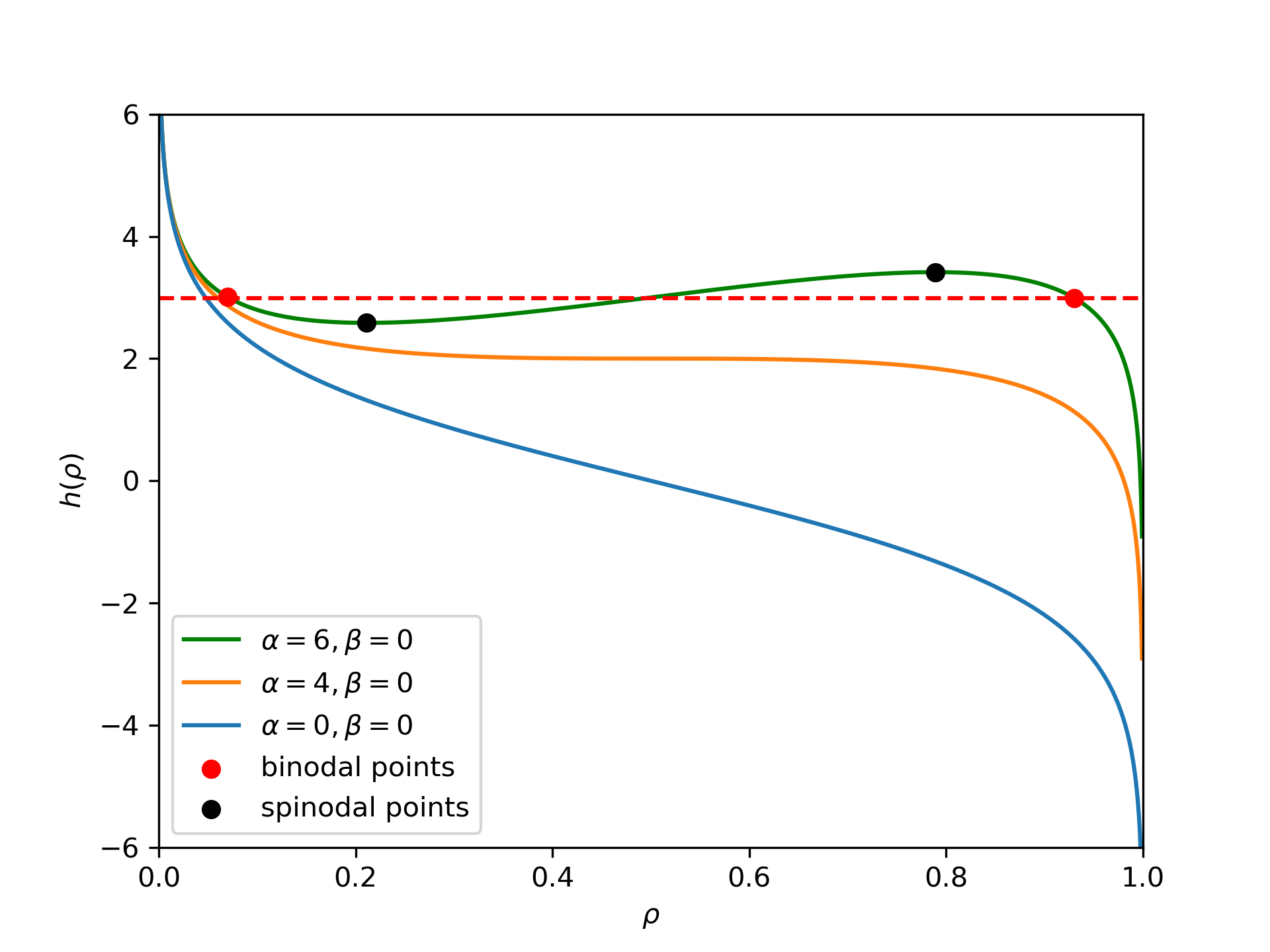}
\caption{Effective Utility vs Density: $h$ vs $\rho$ for different $\alpha$. The black points are the spinodal points ($\rho_{s1} = 0.211, h_{s1} = 2.585; \rho_{s2} = 0.789, h_{s2} = 3.415$). The red points are the binodal points ($\rho_{b1} = 0.071, h_{b1} = 3.00; \rho_{b2} = 0.929, h_{b2} = 3.00$).}
\label{fig:Utility_density_zero_beta}
\end{figure}

\begin{figure}[!ht]
\includegraphics[width=0.5\textwidth]{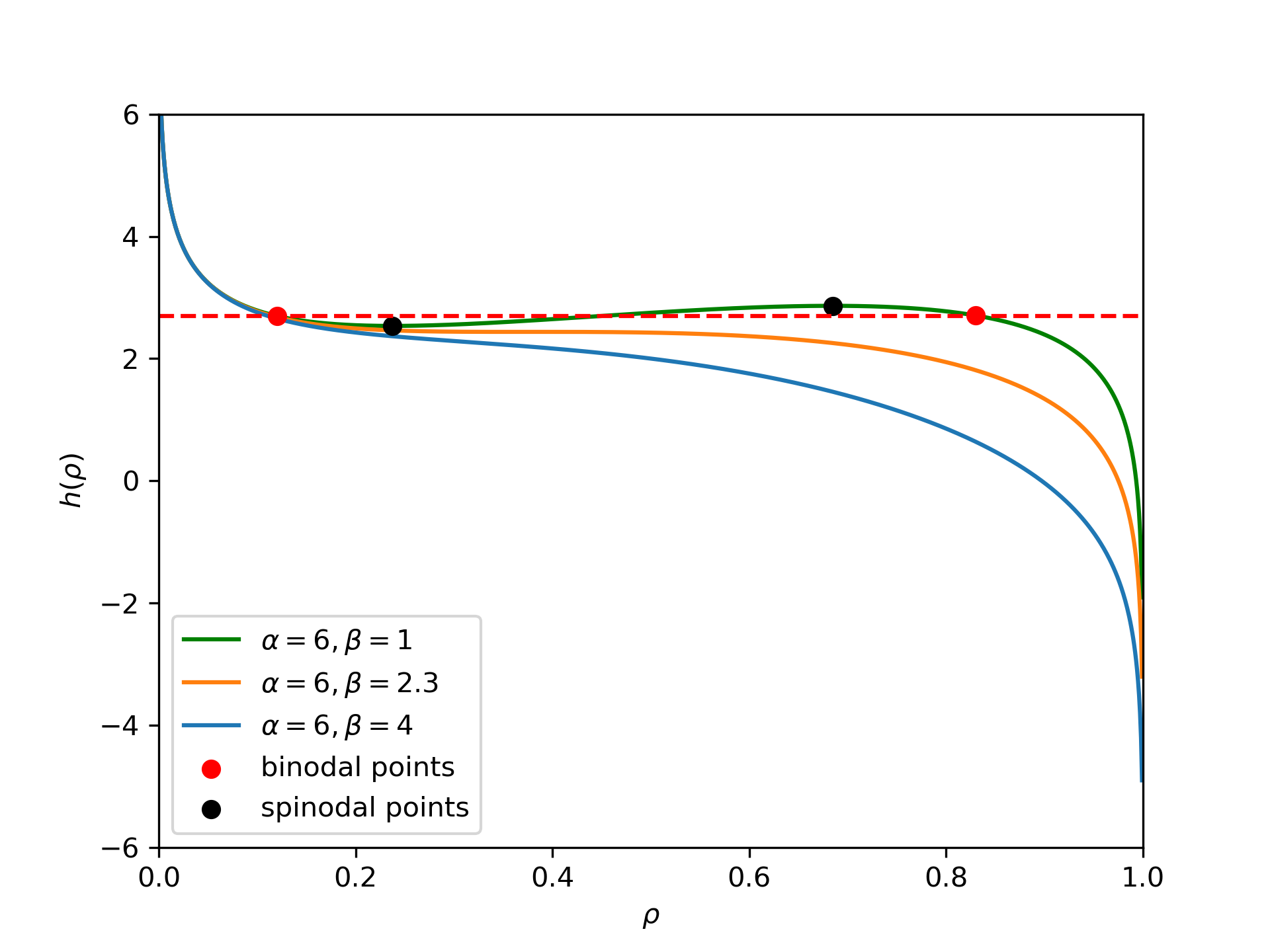}
\caption{Effective Utility vs Density: $h$ vs $\rho$ for different $\beta$
The red points are binodal points. The black points are the spinodal points ($\rho_{s1}$ = 0.237, $h_{s1}$ = 2.535; $\rho_{s2}$ = 0.685, $h_{s2}$ = 2.864). The red points are the binodal points ($\rho_{b1}$ = 0.117, $h_{b1}$ = 2.708; $\rho_{b2}$ = 0.829, $h_{b2}$ = 2.708).}
\label{fig:Utility_density_nonzero_beta}
\end{figure}

We explore numerically the behavior of $h^*$ as a function of $\rho ^*$ in ~\eqref{eq:utility_equil}, as shown in Fig.~\ref{fig:Utility_density_zero_beta} ( $\beta = 0$, different $\alpha$) and Fig.~\ref{fig:Utility_density_nonzero_beta} ( $\alpha = 6$, different $\beta$). As we can see, these two plots are qualitatively similar. Below a threshold value of $\alpha$ and $\beta$, the utility function is monotonic and has a unique density (blue curve) for a given utility value. Above the threshold, the utility is non-monotonic (green curve) and can have multiple density values for the same utility. The red dotted line shows this. The orange curve is the threshold behavior. \\

Note that whether all the agents remain in a single phase of uniform density dispersed throughout the region or they separate into various clumps is determined by the slope $\partial h/\partial \rho\big\vert_{\rho^*}$, which is the second derivative of $\phi$, $\partial^2 \phi /\partial^2 \rho\big\vert_{\rho^*}$. This behavior is mathematically equivalent to \textit{spinodal decomposition} in thermodynamics, widely studied, for example, in the phase separation of alloys and polymer blends~\cite{cahn1961spinodal, favvas2008spinodal}. \\

In thermodynamics, the phase between the spinodal points (discussed below in more detail) is unstable as it corresponds to increasing the free energy of the system, and hence the single phase splits into two phases of different densities to lower the free energy.  For the same reason, the phases between the spinodal and binodal points are metastable, and the phases at the binodal points are stable. \\

A similar behavior happens here in statistical teleodynamics as well. Here, the goal is to maximize the potential $\phi$ in \eqref{eq:schelling-potential}. Thus, in Fig~\ref{fig:Utility_density_zero_beta}  and Fig.~\ref{fig:Utility_density_nonzero_beta}, we observe that for $\alpha = 0$ (blue curve) and $\alpha = 4$ (orange curve), $\partial h/\partial \rho \le 0$ (i.e. negative slope; recall that   $\partial h/\partial \rho $ = $\partial^2 \phi /\partial^2 \rho $). In such a parameter regime, phase separation does not occur. However, for higher values of $\alpha$, regions with $\partial h/\partial \rho>0$ (i.e., positive slope) phase separation develops. 
\\

\begin{figure}[]
\includegraphics[width=0.5\textwidth]{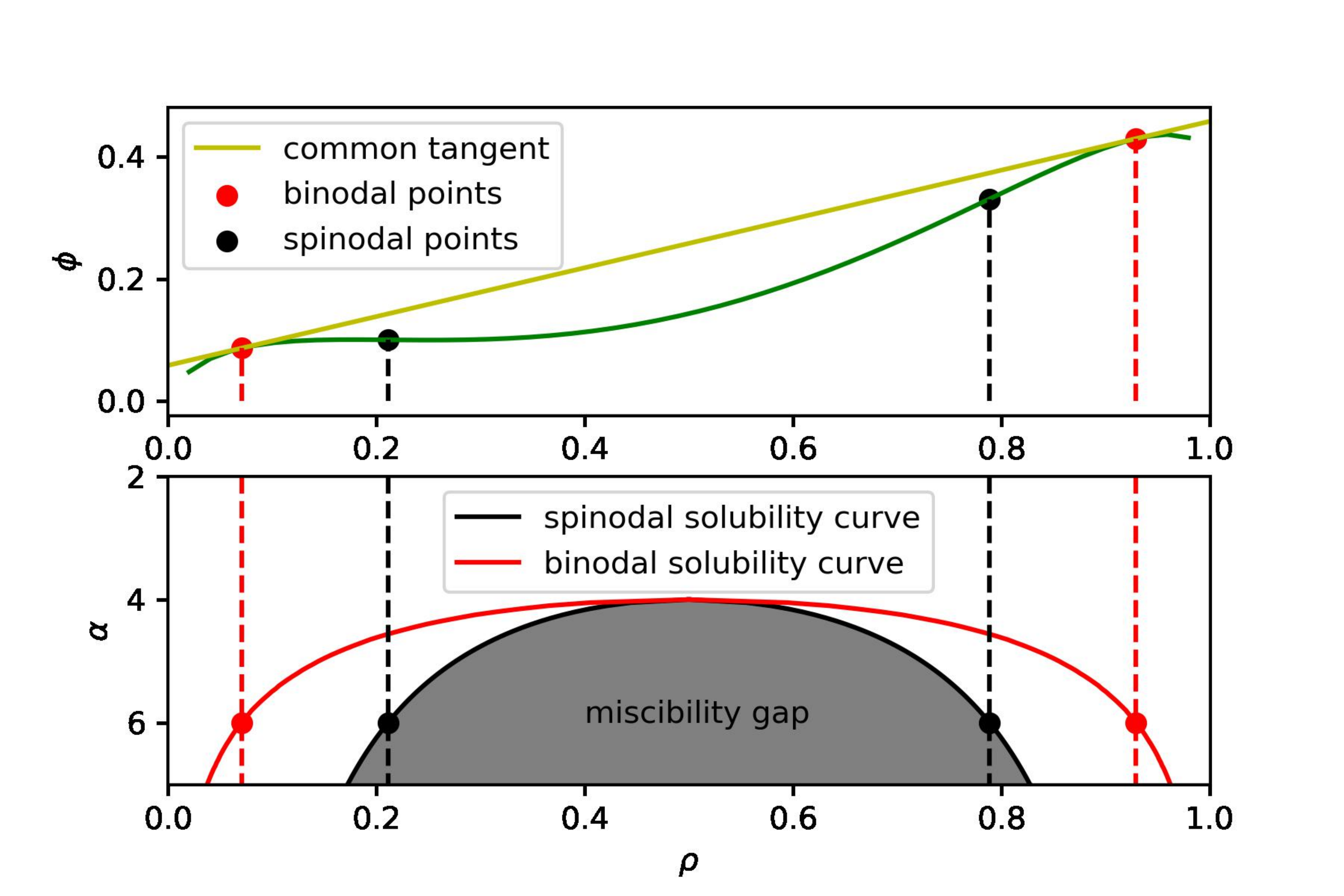}
\caption{Game potential ($\phi$) curve and the spinodal and binodal points. For $\alpha =6, \beta = 0$, the spinodal densities are 0.211 and 0.789; the binodal densities are 0.071 and 0.929.}
\label{fig:SpinodalBinodal}
\end{figure}

We understand this better from Fig.~\ref{fig:SpinodalBinodal}. The top part of this figure displays the potential ($\phi$) vs the density ($\rho$) curve (in green) for $\alpha =6$, $\beta =0$. 
The equation plotted is \\

$\phi = \alpha \frac{\rho^2}{2} - \beta \frac{\rho^3}{3}- \rho \ln{\rho}- (1-\rho) \ln{(1-\rho)}-2.6 \rho$ \\

The linear term $2.6 \rho$ is subtracted from the actual potential function as a way of re-scaling to highlight the \textit{double-hump} nature of the $\phi - \rho$ curve. This subtraction is done purely for illustrative purposes only, as this double-hump otherwise is not so visible in the scale of the figure. In all our calculations and simulations, this subtraction is not needed and hence is not done. \\

The spinodal points are shown as black dots, where  $\partial h/\partial \rho\big\vert_{\rho^*}$ = $\partial^2 \phi /\partial^2 \rho\big\vert_{\rho^*} = 0$. The corresponding spinodal points are also shown in Fig.~\ref{fig:Utility_density_zero_beta} as black dots on the green curve ($\alpha = 6$, $\beta = 0$). Fig.~\ref{fig:SpinodalBinodal} also shows the binodal points (in red, connected by the common tangent line), where  $\partial h/\partial \rho\big\vert_{\rho^*}$ = $\partial^2 \phi /\partial^2 \rho\big\vert_{\rho^*} < 0$. The corresponding binodal points are seen in Fig.~\ref{fig:Utility_density_zero_beta} as red points connected by the red dotted line. As we see, the two binodal points enjoy the same utility (3.00), which is the arbitrage equilibrium.\\

The bottom part of Fig.~\ref{fig:SpinodalBinodal} shows the loci of binodal points (red curve) and of spinodal points (black curve) for different values of $\alpha$ ($\beta = 0$). As $\alpha$ changes, the binodal and spinodal points change, and for $\alpha > 4$ ($\beta = 0$) they disappear. Within the spinodal region, shown in dark grey and known as the miscibility gap in thermodynamics, a single-phase of uniform density is unstable and would split into two-phases of different densities. The reason is the potential $\phi$ of a large mussel clump here is less than the sum of the two potentials of the low-density clump and the high-density clump at the binodal points. We see this geometrically from the common tangent line connecting the binodal points to be above the single-phase green curve between the spinodal points. Agents in such regions will be self-driven towards the high-density binodal point to increase their utility. So, $\phi$ increases and the system splits into two clumps of different densities. Thus, for the green curve in Fig.~\ref{fig:Utility_density_zero_beta}, a self-organized, utility-driven, stable phase separation occurs spontaneously at the binodal points (red dotted line) at the arbitrage equilibrium. While the miscibility gap is unstable, the region immediately outside of it, between the black and red curves, is metastable. Beyond the red curve, one has stable single phase of uniform density - no phase separation here.  \\

In summary, for high values of $\alpha$ (e.g., green curve in Fig.~\ref{fig:Utility_density_zero_beta}), combined with average densities in the miscibility gap, we observe the spontaneous emergence of two phases, high and low density clumps of mussels, at arbitrage equilibrium, biologically-driven by the self-actuated pursuit of maximum utility by the mussels. \\

Intuitively, in the high-density phase, the mussels derive so much more benefit from the affinity term (due to high $\alpha$) that it more than offsets the disutilities due to congestion and competition, thereby yielding a high effective utility. Similarly, in the low-density phase, the benefits of reduced congestion and lower competition combined with increased option benefit more than compensates for the loss of utility from the affinity term. Thus, every agent enjoys the same effective utility $h^*$ in one phase or the other at equilibrium. This causes equilibrium because, as noted, there is no more arbitrage incentive left for the mussels to switch neighborhoods. Again, we wish to remind the reader that our view is that this process occurs instinctively for the mussels, not because of a deliberate rational thought process. \\

As noted, this analysis is mathematically equivalent to spinodal decomposition in statistical thermodynamics, with an important difference. In statistical thermodynamics, agents try to \textit{minimize} their chemical potentials and the free energy of the system. Here, in statistical teleodynamics, the agents try to \textit{maximize} their utilities ($h_i$) and the game theoretic potential ($\phi$). In thermodynamics, chemical potentials are equal at the phase equilibrium. In teleodynamics, the utilities are equal at the arbitrage equilibrium. The parallel is striking, but not surprising, because, as Venkatasubramanian has shown~\cite{venkat2017book, venkat2022unified}, statistical teleodynamics is the generalization of statistical thermodynamics for goal-driven agents.\\

\begin{figure}[!ht]
    \centering
    \includegraphics[width=\linewidth]{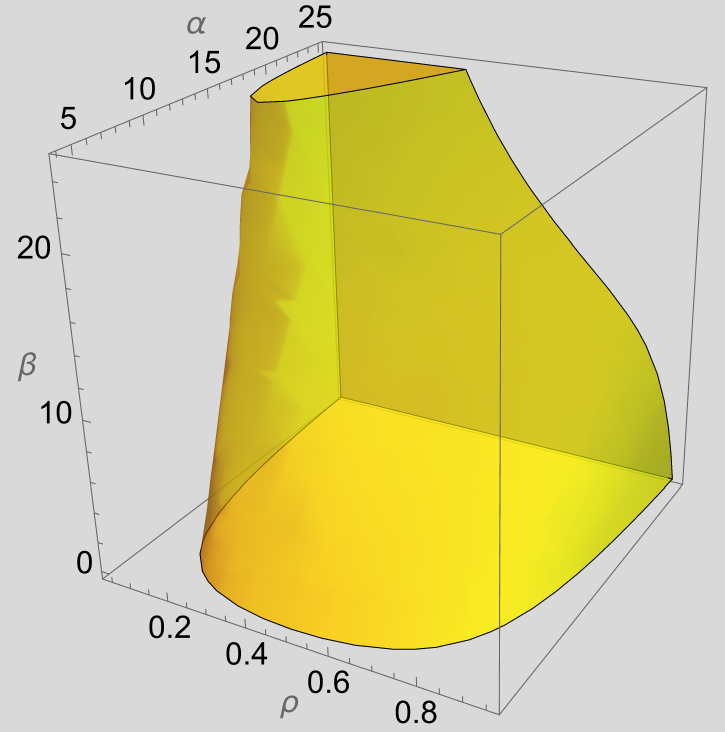}
    \caption{{\bf  Phase separation region at the arbitrage equilibrium}}
    \label{fig:phase_separation}
\end{figure}

\begin{figure}[!ht]
    \centering
    \includegraphics[width=0.5\textwidth]{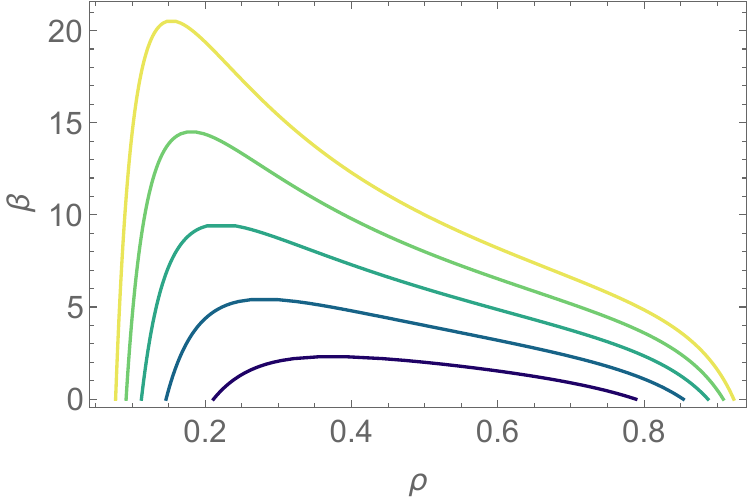}
    \caption{{\bf  Constant $\alpha$ (vertical) 2-D slices of the yellow region}}
    \label{fig:phase_separation_alpha_slice}
\end{figure}

\begin{figure}[!ht]
    \centering
    \includegraphics[width=\linewidth]{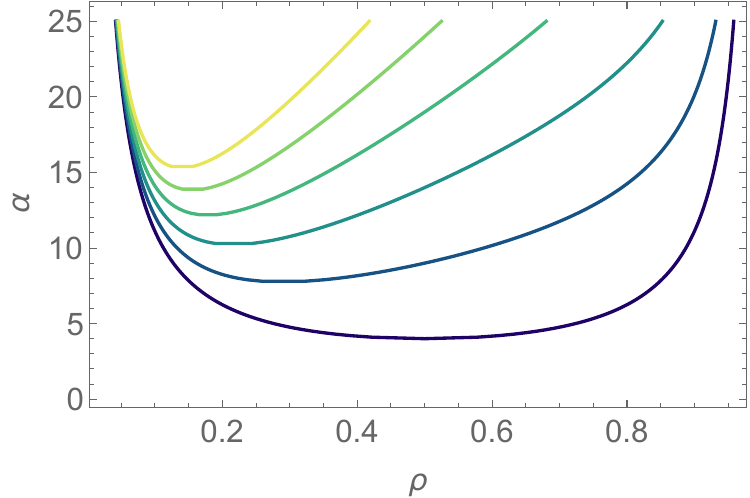}
    \caption{{\bf  Constant $\beta$ (horizontal) 2-D slices of the yellow region}}
    \label{fig:phase_separation_beta_slice}
\end{figure}

In Fig.~\ref{fig:phase_separation}, we show the region (shaded in yellow) within which the phase separation occurs at the arbitrage equilibrium for values of the average density $\rho_0$, $\alpha$, and $\beta$ within the region. In Fig.~\ref{fig:phase_separation_alpha_slice} and Fig.~\ref{fig:phase_separation_beta_slice}, we show the 2-D slices of the yellow region of spontaneous phase separation. For a given value of $\alpha$, $\beta$, and $\rho_0$, they show the loci of the two densities (i.e., the low and high density clumps) of the corresponding equilibrium states of the mussels.

\section{Agent-based simulation results}

We tested our model using an agent-based simulation on a $300\times 300$ lattice (90,000 cells total), for  $\alpha = 6$, $\beta = 0$, and for different $N$ ($N$ = 22,500, 45,000, 55,000). For the details of the simulations, the reader is referred to the Methods section. \\
\begin{figure*}[!ht]
\centering
\includegraphics[scale=0.4]{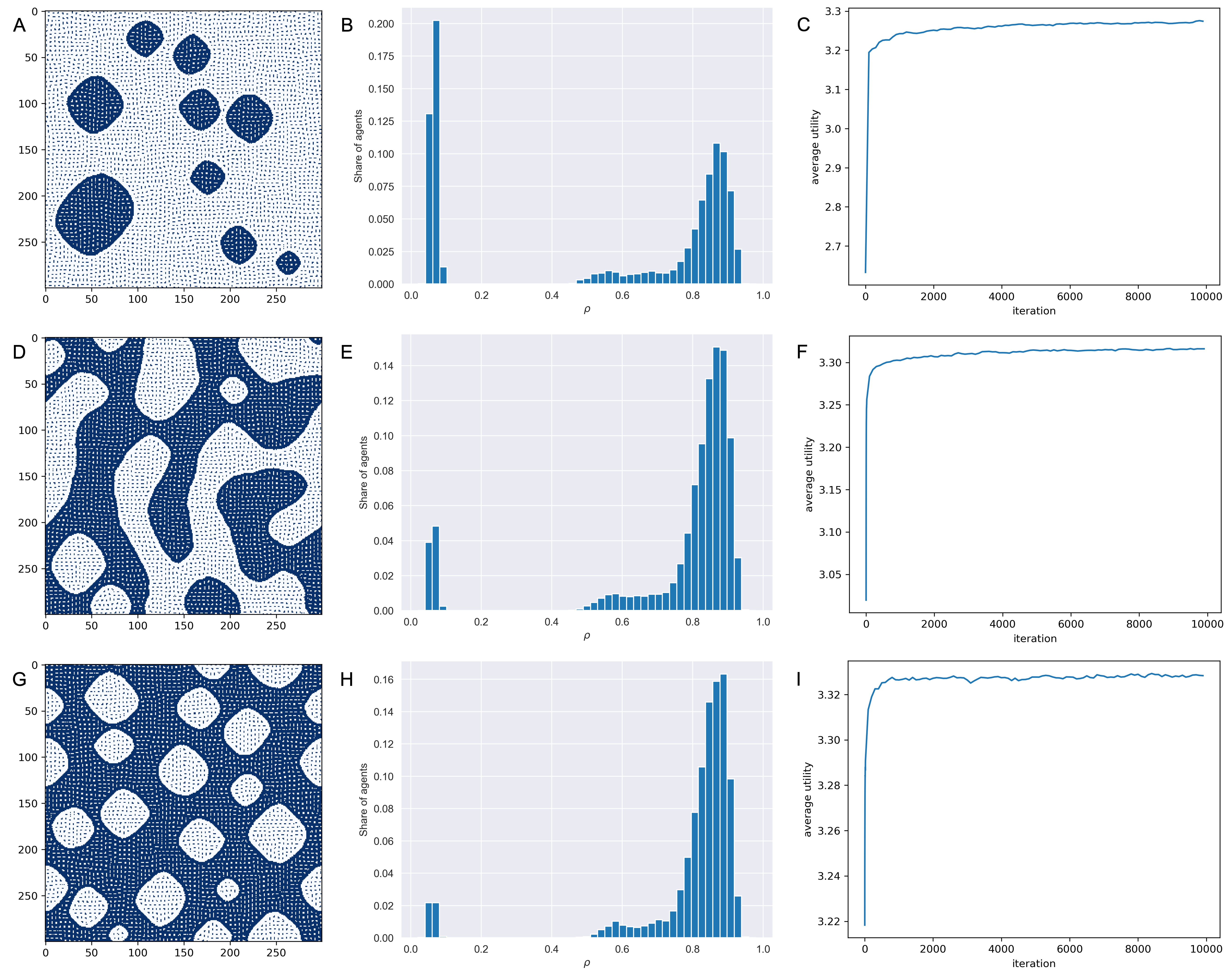}
\caption{Equilibrium patterns for different average densities, $\rho_0$, at the end of 10,000 iterations, for $\alpha = 6,$ $\beta = 0$.
(A), (B), and (C) represent, respectively, the sparsely distributed dots pattern (I) obtained for 22,500 agents ($\rho_0 = 0.25$), the corresponding density distribution at the end of 10,000 iterations, and the average utility evolution over the iterations. Similar results are shown for the labyrinthine pattern (II) in (D), (E), and (F) for 45,000 agents ($\rho_0 = 0.5 $), and for the "gapped" pattern (III) in (G), (H), and (I) for 55,000 agents ($\rho_0 =0.61 $).} 
\label{fig:MusselPatterns_3X3}
\end{figure*}

In our simulations, we observe (Fig.~\ref{fig:MusselPatterns_3X3}) the three basic types of patterns, or "macroscopic" states, namely, (I) sparsely distributed dots, (II) labyrinthine or worm-like structures, and (III) "gapped" patterns  that are seen empirically in mussel beds~\cite{liu2013CahnHilliard} for different mussel densities. As one might expect, the size of the interaction neighborhood (see the Methods section) plays a role in determining the specific details, i.e., the "microscopic" features, of these "macroscopic" states. That is, for example, the detailed "microscopic" features of the labyrinthine or worm-like structures might look different for different neighborhood sizes, but their "macroscopic" state would remain as labyrinthine. Thus, the basic "macroscopic" states are found to be robust. The  "macroscopic" state transitions appear like phase transitions, moving from category I to II to III, as the density increases. \\

We believe these "macroscopic" and "microscopic" characteristics reflect the structure of the phase-space landscape of $\phi$. Thus, as the mussels move around in the physical space, the system wanders around in the phase space landscape, settling into one state or the other. Since a "macroscopic" state could be achieved via many different "microscopic" states - what is known as \emph{multiplicity} in statistical mechanics - one gets different "microscopic" outcomes in different simulation runs, while the "macroscopic" outcome remains the same and robust. The "macroscopic" states are the attractors seen in many nonlinear dynamical systems. \\

The corresponding density histograms and the evolution of average utility as a function of iteration are also shown in Fig.~\ref{fig:MusselPatterns_3X3}. For this configuration (i.e., $\alpha = 6$, $\beta = 0$), the spinodal densities (the black points in Fig.~\ref{fig:Utility_density_zero_beta} and Fig.~\ref{fig:SpinodalBinodal}) are 0.211 and 0.789, and the binodal densities (the red points in Fig.~\ref{fig:Utility_density_zero_beta} and Fig.~\ref{fig:SpinodalBinodal}) are 0.071 and 0.929. \\

A word of caution as we discuss the results. We appreciate that mussels are not molecules, and biology is not physics. Therefore, we do not expect our analytical model, and the associated simulations, to capture all the nuances and the complexities of the real-world patterns in mussel beds. \\

As predicted by our theory, the density histograms show (Fig.~\ref{fig:MusselPatterns_3X3}, B, E, H) spinodal decomposition for all three agent populations ($N$ = 22,500, 45,000, 55,000). That is, there are two phases, one with low-density clumps and the other with high-density clumps. These two densities are around the binodal densities predicted by the theory. While the theory predicts two sharp binodal density values ($\rho_1$ = 0.071 and $\rho_2$ = 0.929), it would be hard to see such precise outcomes in the simulation for one main reason. The theoretical predictions are based on concepts from statistical mechanics, which only work well for extremely large number of agents (such as the Avogadro number of molecules, $10^{23}$). This is when the statistical estimates and outcomes are extremely precise, as in the case of, for example, alloys in materials science. In our simulation, we have only 22,500 - 55,500 agents. So, the statistics are not that precise. Therefore, one should expect to see a distribution of values instead of singular peaks. That is what we observe in our simulations. \\

We also observe that the distributions around the low binodal density are narrower whereas they are broader at the high binodal density. The reason is the following. As we see from Fig.~\ref{fig:Utility_density_zero_beta}, an individual mussel reaches its maximum utility at the upper spinodal point at the spinodal density of 0.789, whereas the whole mussel bed reaches its maximum potential $\phi$ (and hence the arbitrage equilibrium) at the binodal density of 0.929 as seen from Fig.~\ref{fig:SpinodalBinodal}. Thus, in the high density clusters, there is constant jockeying by the individual mussels to reach the upper spinodal point (density = 0.789) of higher individual utility, while the competition from the other mussels to reach the same state drives the mussel bed away from the spinodal point to the binodal point (density = 0.929). Therefore, the mussels are moving around mostly between these two points, the spinodal density of 0.789 and the binodal density of 0.929, with a weighted-average density of about 0.85 (see Table~\ref{tab:summary}) right in the middle. \\

We also observe in Fig.~\ref{fig:MusselPatterns_3X3} (C, F, I) that the average utility improves as the simulation proceeds, as the mussels maneuver around to increase their effective utilities, and then finally settles and fluctuates around the arbitrage equilibrium value.  \\

The key statistics are summarized in Table~\ref{tab:summary}. The spinodal and binodal densities are the same for the three different cases of $N$, because $\alpha = 6, \beta =0$ for all the cases (see Fig~\ref{fig:Utility_density_zero_beta}, green curve). We also find the average utility of Phase-1 to be almost the same as that of the corresponding Phase-2, as predicted by the theory. Thus, we see that a vast majority (86-90\%) of the mussels are in their arbitrage equilibrium states, either in Phase-1 or Phase-2.

\begin{table*}[]
  \centering
  \caption{Summary of key metrics}
    \label{tab:summary}
    \begin{adjustbox}{width=\linewidth,center}
  \begin{tabular}{c c c c c c c c c c c c}
    \toprule
    \multirow{2}{1cm}{Category} & \multirow{2}{1cm}{N} & \multicolumn{2}{c}{Spinodal densities}  & \multicolumn{2}{c}{Binodal densities} & \multicolumn{3}{c}{Phase-1} & \multicolumn{3}{c}{Phase-2}  \\ 
    
        {}                     & {}                      &   {$\rho_{s1}$} & $\rho_{s2}$      & $\rho_{b1}$ & $\rho_{b2}$       & \multirow{2}{2cm}{Share of agents($\%$)} & Avg. density & Avg. utility   & \multirow{2}{2cm}{Share of agents($\%$)} & Avg. density & Avg. utility   \\

        {}   & {} & {}   & {} &{}   & {} &{}   & {} &{}   & {} &{}   & {} \\
        \hline
        {Sparsely distributed dots}   & {22,500} & {0.211}   & {0.789} &{0.071}   & {0.929} &{33.28}   & {0.052} &{3.217}   & {52.63} &{0.855}   & {3.330} \\
        {Labyrinthine}   & {45,000} & {0.211}   & {0.789} &{0.071}   & {0.929} &{8.71}   & {0.051} &{3.234}   & {79.85} &{0.850}   & {3.338} \\
        {Gapped}   & {55,000} & {0.211}   & {0.789} &{0.071}   & {0.929} &{4.32}   & {0.050} &{3.250}   & {85.42} &{0.848}   & {3.341} \\

    \hline
  \end{tabular}
\end{adjustbox}
\end{table*}

\section{Stability of the Arbitrage Equilibrium}

We can determine the stability of this equilibrium by performing a Lyapunov stability analysis \cite{venkat2015howmuch, venkat2017book}. A Lyapunov function $V$ is a continuously differentiable function that
takes positive values everywhere except at the equilibrium point (i.e., $V$ is positive definite), and decreases (or is nonincreasing) along every
trajectory traversed by the dynamical system ($\dot{V}$ is negative definite or negative semidefinite). A dynamical system is locally stable at equilibrium if $\dot{V}$ is negative semidefinite and is asymptotically stable if $\dot{V}$ is negative definite.\\

Following Venkatasubramanian~\cite{venkat2017book}, we identify our Lyapunov function $V(\boldsymbol{\rho})$

\begin{eqnarray}
V(\boldsymbol{\rho}) = \phi^*(\boldsymbol{\rho}) - \phi(\boldsymbol{\rho}) \label{eq:lyapunov}
\end{eqnarray}

where $\phi^*$ is the potential at the arbitrage equilibrium (AE) (recall that $\phi^*$ is at its maximum at AE) and $\phi(\boldsymbol{\rho})$ is the potential at any other state. Note that $V(\boldsymbol{\rho})$ has the desirable properties we seek: (i) $V(\boldsymbol{\rho^*})$ = 0 at AE and $V(\boldsymbol{\rho})$ $>$ 0 elsewhere, i.e., $V(\boldsymbol{\rho})$ is positive definite; (ii) since $\phi(\boldsymbol{\rho})$ increases as it approaches the maximum, $V(\boldsymbol{\rho})$ decreases with time, and hence it is easy to see that $\dot{V}$ is negative definite. Therefore, the arbitrage equilibrium is not only stable but also \textit{asymptotically stable}. \\

\begin{figure*}[!ht]
    \centering
    \includegraphics[width=\linewidth]{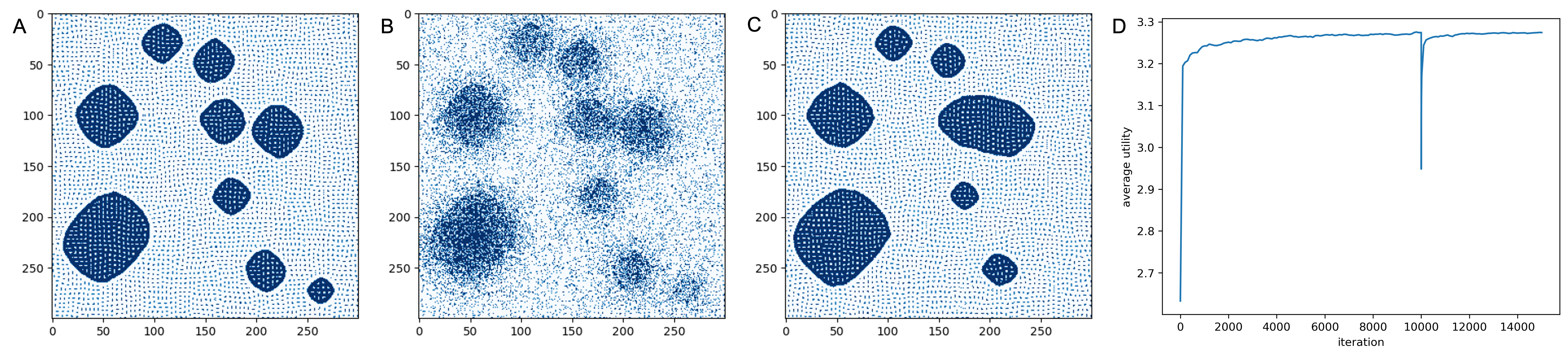}
    \caption{{\bf  Stability analysis. (A) Equilibrium configuration of agents at the end of 10,000 iterations (B) disturbed configuration at 10,001t$^\text{th}$ iteration (C) equilibrium configuration at the end of 15,000 iteration  (D) evolution of average utility over iterations. The sharp decrease in the average utility at 10,001$^\text{th}$ is due to the disturbance introduced at 10,001$^\text{th}$ iteration.}}
    \label{fig:Stability_analysis_agent_locations}
\end{figure*}

Our simulation results confirm this theoretical prediction (see Fig.~\ref{fig:Stability_analysis_agent_locations}). We show the stability results for the configuration of Fig.~\ref{fig:MusselPatterns_3X3}-A, as an example. After the mussels population reached equilibrium (10,000 iterations), we disturbed the equilibrium state by randomly changing the positions of the mussels. As a result, the average utility of the population goes down as seen from the sharp drop at the 10,001th iteration in Fig.~\ref{fig:Stability_analysis_agent_locations}-D. Fig.~\ref{fig:Stability_analysis_agent_locations}-B shows the disturbed state of the mussel clumps. The simulation is then continued from this new disturbed far-from-equilibrium state. As we see from Fig.~\ref{fig:Stability_analysis_agent_locations}-C, the mussels population recovers quickly to reach the original category-I "macroscopic" state even though some of the microscopic features are different this time. The reader might have noticed that two small clumps in Fig.~\ref{fig:MusselPatterns_3X3}-A have merged become one larger clump in two different locations in Fig.~\ref{fig:MusselPatterns_3X3}-C. This phenomenon is called Ostwald Ripening in materials science. We also notice that the average utility is back to its old level. \\

This analysis shows that the arbitrage equilibrium region is not only stable, but asymptotically stable. That is, the mussel beds are resilient and self-healing. Given the speed of the recovery, it could possibly be exponentially stable but we have not proved this analytically here. It is interesting to observe that this result is similar to that of the dynamics of the income-game~\cite{venkat2015howmuch, venkat2017book} and the dynamics of birds flocking~\cite{venkat2022garuds}. 

\section{Conclusions}

A number of physics- and chemistry-based mechanisms, such as the scale-dependent feedback~\cite{koppel2005scale} and Cahn-Hilliard inspired model~\cite{liu2013CahnHilliard}, have been proposed over the years for the emergence of spontaneous patterns in mussel beds. Here, we propose a different perspective that utilizes concepts and techniques from economics and game theory. Since mussels are biological agents driven by the purpose to survive and thrive in challenging environments, modeling this survival goal explicitly is central in our theory.\\

We thus adapt the concept of utility from economics and game theory to model this survival-fitness objective, which the mussels try to maximize instinctively by their dynamic behavior. They maneuver around in their turbulent environments, exploiting local arbitrage opportunities to maximize their survival-fitness. This self-organizing dynamical behavior eventually leads to an asymptotically stable arbitrage equilibrium and pattern formation, as we showed above. \\

While it is well-known that the dynamics of inanimate matter is determined by invariants (i.e., constants of motion such as total energy and momentum), it is surprising to find that the dynamics of living systems, such as mussel populations, could also have an invariant, namely, the effective utility. As we saw from the stability analysis of our model and the simulations, once the mussel population achieves the arbitrage equilibrium it tends to stay there despite disturbances. However, the role of invariance here is different from its role in dynamics. The constants of motion such as energy and momentum are conserved, but effective utility is not. The role of this invariance is more like that of set-point tracking and disturbance rejection in feedback control systems \cite{seborg2016process, aastrom2021feedback}. The system, i.e., the mussels population, adjusts itself dynamically and continually, in a distributed feedback control-like manner, to maintain its overall maximum effective utility. \\

It is important to emphasize, however, that this control action is decentralized as opposed to the typical centralized control system in many engineering applications. The agents individually self-organize, adapt, and dynamically course-correct to offset the negative impact on their effective utilities by other agents or other external sources of disturbance. The population as a whole stochastically evolves towards the stable basin of attraction in the phase space in a self-organized and distributed-control fashion.\\

We are keenly aware, of course, of the simplifications we have made to make the analysis analytically tractable.  We realize that our model agents are not real mussels, and our model and simulations are not real biological systems. They are stylized ideal systems, formulated in the spirit of similar ideal systems in statistical mechanics, such as the ideal gas and the Ising model. Despite such an ideal approximation, our results nevertheless suggest intriguing possibilities for real biological and ecological entities that need to be explored further.\\

We suggest that this pursuit of maximum utility or survival-fitness is a universal self-organizing mechanism. In the case of mussels, the incessant search for improving the survival fitness occurs in the physical space of the sea, with the mussels trying to move to a better location in this environment. In biology, the search for improving one's fitness occurs in the design space of genetic features. Here, the mutation and crossover operations facilitate the movements in this features space, such that an agent improves itself genetically to increase its utility, i.e., the survival-fitness. In economics, agents search in the products/services space, so that they can offer better products/services to improve their economic-survival-fitness in a competitive market place. Thus, our mechanism is essentially the same as Adam Smith's \textit{invisible hand}. Every agent is pursuing its own self-interest, to increase its own $h_i$, and a stable collective order spontaneously emerges via self-organization as we show.\\

This utility-focused mathematical framework has been demonstrated for other dynamical systems in biology \cite{venkat2022unified}, ecology~\cite{venkat2022garuds}, economics~\cite{venkat2015howmuch, venkat2017book, kanbur2020occupational}, and sociology \cite{venkat2022unified} to predict emergent phenomena via self-organization. Venkatasubramanian et al.~\cite{venkat2015howmuch} showed that the emergence of the exponential energy (i.e., Boltzmann) distribution for gas molecules can be modeled by the effective utility 

\begin{equation}
h_{i}= -\beta E_i - \ln N_i.
\label{eq:Boltzmann_utility}
\end{equation}

Similarly,  Venkatasubramanian et al. showed~\cite{venkat2022unified} for biological systems, the benefit-cost trade-off in effective utility $h_i$ for bacterial chemotaxis can be modeled by

\begin{equation}
h_{i}= \alpha c_i - \ln N_i
\end{equation}

where the first term is the benefit derived from a resource ($c_i$, $\alpha >0$) and the second is the cost of competition as modeled in Eqn.~\ref{eq:Utility_N_i}.\\

The emergence of ant craters can be modeled by~\cite{venkat2022unified} 

\begin{equation}
    h_i = b - \frac{\omega r_i^a}{a} - \ln N_i. 
\end{equation}

where the first term ($b$) is the benefit of having a nest for an ant, the second term is the cost of work done in carrying the sand grains out to build the nest, and the last term is again the cost of competition as before. \\

In another study in ecology~\cite{venkat2022garuds}, Sivaram and Venkatasubramanian showed how the flocking behavior of bird-like agents can be modeled by

\begin{equation}
    h_i = \alpha N_i - \beta N_i^2+ \gamma N_i l_i - \ln N_i 
    \label{eq:garuds_utility}
\end{equation}

where the first term is the benefit of aggregation, the second is the cost of congestion, the third is the benefit of alignment, and the last is again the cost of competition.\\

Finally, in economics, the emergence of an income distribution can be modeled by  

\begin{equation}
h_{i}= \alpha \ln S_i - \beta \left(\ln S_i\right)^2 - \ln N_i.
\label{eq:income_utility}
\end{equation}

where the first term is the benefit of income, the second is the cost of work, and the last is again the cost of competition.\\

By comparing the equations \ref{eq:Boltzmann_utility} through \ref{eq:income_utility} with Eq.~\ref{eq:Utility_N_i} 

\begin{eqnarray*}
    h_i = \alpha N_i  - \beta N_i^2 + \ln(M - N_i) - \ln N_i
\end{eqnarray*}

for the mussels, we observe a certain universality in the structure of the effective utility functions in different domains. They are all based on benefit-cost trade-offs, except that the actual nature of the benefits and costs depend on the details of the specific domains as one would expect. Thus, we see that the same conceptual and mathematical framework is able to predict and explain the emergence of spontaneous order via self-organization to reach arbitrage equilibrium in dynamical systems in physics, biology, ecology, sociology, and economics. We find this universality appealing and reassuring. \\

What we have here is for ideal systems, of course, similar to the ideal gas or the Ising model in thermodynamics. Just as real gases and liquids don't behave exactly like their ideal versions in statistical thermodynamics, we don't expect real biological systems (or economic or ecological systems) to behave like their ideal counterparts in statistical teleodynamics. Nevertheless, just as the ideal versions continue to serve as useful starting points in statistical thermodynamics, we expect our model systems to be similarly helpful in statistical teleodynamics of active matter.  The next steps would involve examining and learning how real-world biological systems behave compared to their ideal versions. This would, of course, necessitate several modifications to the ideal models. 

\section{Methods}

The agent-based simulation was performed using Python. We distributed agents on a 2-D $ 300 \times 300$ grid with 90,000 cells. Three simulation studies are reported in this paper - with 22,500 agents, 45,000 agents, and 55,000 agents. For each case, initially, the agents were randomly distributed on the grid with each agent occupying one cell. \\

The dynamical evolution of the system is determined by two neighborhoods around an agent $i$. One is the local neighborhood of \emph{interaction}, which is an area with 49 cells surrounding agent $i$ (including the cell $i$ is occupying). The other is the \emph{exploration} neighborhood (which is larger than the interaction neighborhood and contains it) within which an agent $i$ can explore and move to another cell to improve its utility $h_i$. The exploration neighborhood has 1680 cells. The neighborhood sizes are parameters that can be varied to balance the need to allow for complex patterns to emerge at arbitrage equilibrium and the need to accomplish this in a reasonable amount of computational time. We found that our combination (49 and 1680) accomplishes this well. \\

The density of agents at any cell is defined as the ratio of the number of agents in the interaction neighborhood to the total number of cells in the neighborhood. At each iteration, every agent is given an opportunity to move to a vacant cell in the exploration neighborhood where it would have a higher utility than its current cell. If the agent does not find a vacant cell, it chooses to stay at its current location. After an agent moves, its utility and its neighbors' density and utility are updated. The simulations were carried out for 10,000 iterations by which time the system typically reached the arbitrage equilibrium. \\

\vskip6pt

\subsection*{Acknowledgements} 
This work was supported in part by the Center for the Management of Systemic Risk (CMSR), Columbia University. The authors acknowledge the assistance of Leo Goldman with the agent-based modeling software.

\bibliographystyle{apsrev4-2}
%

\end{document}